\documentclass[oneocolumn,preprintnumbers,amsmath,amssymb]{revtex4}

\usepackage{amsmath,latexsym,amssymb,amsfonts}
\usepackage[dvips]{graphicx,color}
\usepackage{xcolor}
\usepackage{ulem}
\usepackage{bm}
\usepackage{gensymb}

\begin{document}


\title{\textbf{Novel phenomena of the Hartle-Hawking wave function}}

\author{
Subeom Kang$^{a}$\footnote{{\tt subeom527@gmail.com}},
Wan-il Park$^{b}$\footnote{{\tt wipark@jbnu.ac.kr}}
and
Dong-han Yeom$^{c,d}$\footnote{{\tt innocent.yeom@gmail.com}}\footnote{All authors contributed equally.}
}

\affiliation{
$^{a}$Department of Physics, Sungkyunkwan University, Suwon 16419, Republic of Korea\\
$^{b}$Division of Science Education and Institute of Fusion Science, Jeonbuk National University, Jeonju 54896, Republic of Korea\\
$^{c}$Department of Physics Education, Pusan National University, Busan 46241, Republic of Korea\\
$^{d}$Research Center for Dielectric and Advanced Matter Physics, Pusan National University, Busan 46241, Republic of Korea
}

\begin{abstract}
We find a novel phenomenon in the solution to the Wheeler-DeWitt equation by solving numerically the equation assuming $O(4)$-symmetry and imposing the Hartle-Hawking wave function as a boundary condition.
In the slow-roll limit, as expected, the numerical solution gives the most dominant steepest-descent that describes the probability distribution for the initial condition of a universe. 
The probability is consistent with the Euclidean computations, and the overall shape of the wave function is compatible with analytical approximations, although there exist novel differences in the detailed probability computation. Our approach gives an alternative point of view of the no-boundary wave function from the wave function point of view. Possible interpretations and conceptual issues of this wave function are discussed.
\end{abstract}

\maketitle

\newpage

\tableofcontents

\section{Introduction}

Understanding the origin of our universe is an important but unresolved problem in modern theoretical physics. 
Due to the singularity theorem \cite{Hawking:1970zqf}, the beginning of our universe must be a kind of singularity. 
This singularity can be directly resolved by quantizing the spacetime and fields. 
We do not know the fundamental theory of quantum gravity yet, but we can write and solve the wave equation of the quantized spacetime and matter fields using the canonical approach \cite{DeWitt:1967yk}. 
The master equation of the quantum Hamiltonian constraint in this approach is called the \textit{Wheeler-DeWitt equation}. 
The application of the canonical quantum gravity to the cosmological context is known as \textit{quantum cosmology}.

It is well-known that one can write and solve the Wheeler-DeWitt equation, at least in minisuperspace \cite{Vilenkin:1987kf} 
In the presence of a matter field $\phi$, we may assume that the wave function of the universe $\Psi$ is a function (functional) of the metric $a$ and $\phi$.
The physical interpretation of $\Psi$ in this case is not clear, but we can get some ideas of possible interpretations from quantum mechanics.
\begin{itemize}
\item[--] 1. \textit{Wave packet}: If we consider a wave function as a superposition of eigenstates, we obtain the following wave function:
\begin{eqnarray}
\Psi(x,t) = \sum_{n} c_{n} \Psi_{n}(x) e^{- i E_{n} t}, 
\end{eqnarray}
where $\Psi_{n}$ denotes the $n$-th eigenstates and $E_{n}$ is the corresponding energy eigenvalue. 
In this wave packet, one can provide the meaning of the classicality: if a wave function is highly localized, the Ehrenfest theorem shows that the expectation value of an observable satisfies the classical equation of motion.
\item[--] 2. \textit{Eigenstate}: If we only choose one specific eigenstate, e.g., the ground state, we will only focus on $\Psi_{n}(x)$ with a fixed energy eigenvalue. The probability to measure at $x$ is $|\Psi_{n}(x)|^{2}$. In the scattering state, if the wave function is oscillatory, say $e^{\pm ikx}$, it indicates a classical propagation, while if the wave function is exponentially varied, say $e^{\pm \kappa x}$, it indicates a quantum regime. 
Here, the latter can be well approximated by the Wentzel-Kramers-Brillouin (WKB) approximation, and this can provide a tunneling or nucleation probability.
\end{itemize}
In this regard, we should rely on several alternative intuitions from quantum mechanics to understand the meaning of the wave function $\Psi[a,\phi]$. 
There might be several hypotheses:
\begin{itemize}
\item[--] 1. \textit{Wave packet interpretation}: If one interprets the wave function as a superposition of various states, one can define a propagating wave packet \cite{Kiefer:1988tr}. 
From this wave packet, one can read a classical trajectory that is consistent with the Ehrenfest theorem. 
If the wave function is flat along the steepest-descent while its dispersion is bounded, and hence, if the probability is not varied along the path, one can interpret that the trajectory is classical; indeed, it can satisfy the classical equation of motion \cite{Bouhmadi-Lopez:2019kkt}. 
Therefore, one can reasonably recover an arrow of time. 
So, the wave function describes the classical dynamics of the universe (although it can include non-classical effects, e.g., quantum bounces, thanks to the wave nature of the wave function).
\item[--] 2. \textit{Eigenstate interpretation}: If one interprets the wave function as a specific eigenstate, one will first interpret that the wave function has two limits, where the wave function shows exponential behaviors in the classically disallowed domain, while it shows oscillatory behaviors in the classically allowed domain \cite{Vilenkin:1987kf}. 
Hence, at the classically disallowed domain, one will measure a specific $a$ and $\phi$, where \textit{the absolute square of the wave function $|\Psi|^{2}$ is the probability to measure $a$ and $\phi$}. 
Once one measures a specific $a$ and $\phi$, the universe will evolve along the classical path; hence, the measured $a$ and $\phi$ become a set of initial conditions. 
If one selects this interpretation, we cannot see any subsequent classical dynamics from the wave function of the universe; this only provides a probability distribution.
\end{itemize}
Now one can ask which interpretation is true. 
This is partly up to our interests.
If we consider a definite classical boundary condition for the beginning, the second interpretation (wave packet interpretation) is the case, and hence we do not compare with other alternative histories. 
However, if we consider a global probability distribution of the many-world, one may not be able to find a classical trajectory from the steepest-descent; rather, one can obtain a probability distribution as a function of initial conditions.

In this paper, we focus on the Hartle-Hawking wave function \cite{Hartle:1983ai} 
(see also \cite{Halliwell:1989dy}; the recent updates are in \cite{Lehners:2023yrj,Alexander:2020ymj}), which is supposed to be the \textit{ground state} wave function of the universe. Therefore, it is more natural to interpret the second interpretation. 
One can distinguish the classical domain as well as the quantum domain by looking at the oscillatory behavior of the wave function.
In addition, we will go one step further by solving the Wheeler-DeWitt equation numerically. 
Our approach is new in the sense that we study the matter field direction and the metric field direction at the same time using the numerical method, while the previous literature is restricted to only one direction (either the metric direction \cite{Vilenkin:1987kf} or the field direction \cite{Hartle:2008ng} only) after approximations.
Thanks to this numerical approach, one could go beyond the WKB approximation, and one could find a new kind of phenomenon that does not show up in simplified analytic approaches.

This paper is organized as follows. In Sec.~\ref{sec:pre}, we describe how we can realize the Hartle-Hawking wave function for the numerical method. In Sec.~\ref{sec:nov}, we first confirm that, in the slow-roll limit, our numerical results are consistent with previously obtained results, e.g., instanton approaches; secondly, we show that our results show some novel phenomena. 
Finally, in Sec.~\ref{sec:con}, we conclude with proposals of possible future research topics.

\section{\label{sec:pre}Preliminaries}

In this section, we briefly review the Wheeler-DeWitt equation \cite{Kiefer:2007ria} with a minimal content of matter in the universe, and a specific choice of the boundary condition, the Hartle-Hawking wave function \cite{Hartle:1983ai}.
The metric signature was taken to be $(-,+,+,+)$, and Planck units are used throughout the paper.

\subsection{Wheeler-DeWitt equation}

We consider a model of Einstein gravity with a scalar field as the matter content of the universe, defined by the following action:
\begin{eqnarray}
S = \int d^{4}x \sqrt{-g} \left[ \frac{\mathcal{R}}{16\pi} - \frac{1}{2} \left( \partial \phi \right)^{2} - V(\phi) \right],
\end{eqnarray}
where $\mathcal{R}$ is the Ricci scalar, $\phi$ is a scalar field, and $V(\phi)$ is the potential of the scalar field.
Also, we assume the following metric ansatz:
\begin{eqnarray}
ds^{2} = - N^{2}(t) dt^{2} + a^{2}(t) d\Omega_{3}^{2},
\end{eqnarray}
where $N(t)$ is the lapse function, $a(t)$ is the scale factor, and $d\Omega_{3}^{2}$ is the three-sphere. 
In this minisuperspace model, the action with a homogeneous $\phi$ is reduced as follows:
\begin{eqnarray}
S = 2\pi^{2} \int dt N \left[ \frac{3}{8\pi} \left( - \frac{a \dot{a}^{2}}{N^{2}} + a \right) + \frac{1}{2} a^{3} \frac{\dot{\phi}^{2}}{N^{2}} - a^{3} V(\phi) \right].
\end{eqnarray}
From this action with a Lagrangian defined by $S = \int dt L$, the Hamiltonian $H$ of the model is found through the Legendre transformation:
\begin{eqnarray}
H \equiv p_{a} \dot{a} + p_{\phi} \dot{\phi} - L,
\end{eqnarray}
where $p_{a}$ and $p_{\phi}$ are canonical momenta defined as
\begin{eqnarray}
p_{a} &\equiv& \frac{\partial L}{\partial \dot{a}} = - \frac{3\pi a \dot{a}}{2N},\\
p_{\phi} &\equiv& \frac{\partial L}{\partial \dot{\phi}} = \frac{2\pi^{2} a^{3}\dot{\phi}}{N}.
\end{eqnarray}
It leads to
\begin{eqnarray}
H = N \left[ - \frac{p_{a}^{2}}{3\pi a} + \frac{p_{\phi}^{2}}{4\pi^{2} a^{3}} - \frac{3\pi a}{4} + 2\pi^{2} a^{3} V(\phi) \right].
\end{eqnarray}

The quantum Hamiltonian constraint equation, or the Wheeler-DeWitt equation, is obtained by substituting canonical momentum to operators:
\begin{eqnarray}
p_{a} &\rightarrow& \frac{\partial}{i \partial a}, \\
p_{\phi} &\rightarrow& \frac{\partial}{i\partial \phi},
\end{eqnarray}
We use the following form of the Wheeler-DeWitt equation \cite{Kiefer:2007ria}:
\begin{eqnarray}
\hat{H} \psi[a,\phi] = \left[ \frac{1}{3\pi a^{p+1}} \frac{\partial}{\partial a} \left( a^{p} \frac{\partial}{\partial a} \right) - \frac{1}{4\pi^{2} a^{3}} \frac{\partial^{2}}{\partial \phi^{2}} - \frac{3\pi a}{4} + 2\pi^{2} a^{3} V(\phi) \right] \psi[a,\phi] = 0,
\end{eqnarray}
where $p$ is an arbitrary constant coming from the ambiguity of the operator ordering. 
By choosing the Laplace-Beltrami operator, we can set $p = 1$ \cite{Kiefer:2007ria}. 
The Wheeler-DeWitt equation is then simplified as
\begin{eqnarray} \label{eq:simp-WD-eq}
\left[ \frac{\partial^{2}}{\partial \alpha^{2}}  - \frac{3}{4\pi} \frac{\partial^{2}}{\partial \phi^{2}} - \frac{9\pi^{2}}{4} e^{4\alpha} + 6\pi^{3} e^{6\alpha} V(\phi) \right] \psi[\alpha,\phi] = 0,
\end{eqnarray}
where $\alpha \equiv \ln a$. This is the equation that we will solve in this paper.

\subsection{Hartle-Hawking wave function}

Physically meaningful solutions of the Wheeler-DeWitt equation need a boundary condition, e.g., $\psi[\alpha_*, \phi]$, $\psi[\alpha, \phi_{\rm min}]$, and $\psi[\alpha, \phi_{\rm max}]$ with $\alpha_*$, $\phi_{\rm min}$, and $\phi_{\rm max}$ being a specific set of parameters associated with a given model. 
One of the famous choices is the Hartle-Hawking wave function \cite{Hartle:1983ai}. 
In the formal sense, it is described by the Euclidean path integral:
\begin{eqnarray}
\psi[\tilde{g},\tilde{\phi}] = \int \mathcal{D}g \mathcal{D}\phi \;\;e^{-S_{\mathrm{E}}[g,\phi]},
\end{eqnarray}
where we sum over all regular and compact Euclidean geometries that have the boundary values $\tilde{g} = \partial g$ and $\tilde{\phi} = \partial \phi$. 
This is a solution of the Wheeler-DeWitt equation, at least, in the formal sense.

There are technically two ways to evaluate this wave function. 
The first approach is to compute the path integral in an approximate way using instantons \cite{Hartle:2008ng}. Instantons must be regular and compact, and hence this removes the initial boundary; this is the reason why this wave function is called by the no-boundary wave function. However, this does not indicate that there is no beginning of the universe; the universe has a beginning in terms of the Lorentzian time. Therefore, this can be distinguished from the stationary universe.
The second approach is to solve the Wheeler-DeWitt equation and impose the boundary condition on the wave function \cite{Vilenkin:1987kf}. 
Both approaches are consistent in the spirit of the WKB approximation; however, in the second approach, we lose the notion of the Euclidean no-boundary condition. 
Two approaches must be equivalent, but since we need to rely on approximations, these approaches will be technically complementary to each other.

In this work, we take the second approach and approximately solve the Wheeler-DeWitt equation with the extremely slow-roll condition, i.e., $\phi = \mathrm{const}$. 
For sub-horizon scales, i.e., $a < 1/\mathcal{H}$ with $\mathcal{H}(\phi) \equiv \sqrt{8 \pi V(\phi)/3}$, the no-boundary wave function can be approximated by
\begin{eqnarray}\label{eq:nb}
\psi\left[a, \phi \right] \propto \frac{1}{\left( 1 - \mathcal{H}^2 a^{2} \right)^{1/4}} \exp \left[ \frac{\pi}{2 \mathcal{H}^2} \left( 1 - \left( 1 - \mathcal{H}^2 a^{2} \right)^{3/2} \right) \right].
\end{eqnarray}
This approximation is sufficiently good if $a \ll 1/\mathcal{H}$.
Therefore, when we solve the Wheeler-DeWitt equation in a region deep inside the horizon, i.e., $a \ll 1/\mathcal{H}$, we will impose the boundary condition, using the analytic form of Eq.~(\ref{eq:nb}).
It will be a good implementation of the Hartle-Hawking wave function in numerical computations.

Interestingly, the Hartle-Hawking wave function, Eq.~(\ref{eq:nb}), indicates that the complete solution would have its maximum at the horizon, $a = \mathcal{H}^{-1}$ which corresponds to the Wick-rotation surface of de Sitter instantons. 
Also, it is worthwhile to mention that in the classical limit ($a > \mathcal{H}^{-1}$, i.e., super-horizon regime), the wave function is well approximated by the following function:
\begin{eqnarray}\label{eq:nb2}
\psi\left[a, \phi \right] \propto \frac{1}{\left(\mathcal{H}^2 a^{2} - 1 \right)^{1/4}} \exp \left[ \frac{\pi}{2 \mathcal{H}^2} \right] \cos \left[ \frac{\pi}{2 \mathcal{H}^2} \left( \mathcal{H}^2 a^{2} - 1\right)^{3/2} - \frac{\pi}{4}\right].
\end{eqnarray}
This will be a good guideline to compare with numerical computations for checking its consistency.

\subsection{Model and boundary conditions}

Technically, to obtain $\psi[\alpha,\phi]$ using the hyperbolic equation, we need, for example, the following boundary conditions: (1) $\psi[\alpha_{0},\phi]$, (2) $\psi[\alpha,\phi_{a}]$, and (3) $\psi[\alpha,\phi_{b}]$, where $\alpha_{0}$ is a constant and $\phi_{a,b}$ are arbitrary boundary values of the field space. 
For (1), we impose the Hartle-Hawking wave function by imposing Eq.~(\ref{eq:nb}) in the $a \ll \mathcal{H}^{-1}$ limit. 
However, it can be a little bit subtle to provide the boundary condition for (2) and (3).
\begin{figure}
\begin{center}
\includegraphics[scale=0.8]{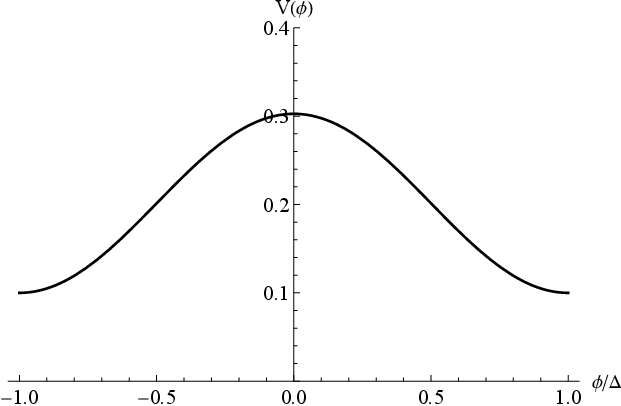}
\caption{\label{fig:potential}The potential $V(\phi)$ for $V_{0} = 0.1$, $m = 1$, and $\Delta = 3$.}
\end{center}
\end{figure}
This problem can be avoided if we can impose the periodic boundary condition, i.e., $\psi[\alpha, \phi_a] = \psi[\alpha, \phi_b]$. 
As an example, we consider a periodic potential model as follows (Fig.~\ref{fig:potential}):
\begin{eqnarray}
V(\phi) = V_{0} + \frac{m^{2}}{\pi^{2}} \left( 1 + \cos \frac{\pi}{\Delta} \phi \right),
\end{eqnarray}
where $V_{0}$ and $m$ are constants and $\Delta \phi$ is the periodicity of the field space. Apparently, this model has the local minimum at $\phi = -\Delta$ and $\phi = +\Delta$.

\section{\label{sec:nov}Novel phenomena of the Hartle-Hawking wave function}

In this section, we report the numerical results of the solution to the Wheeler-DeWitt equation. We focus on the slow-roll limit and compare it with analytic results.

\subsection{Wave function}

\begin{figure}
\begin{center}
\includegraphics[scale=1]{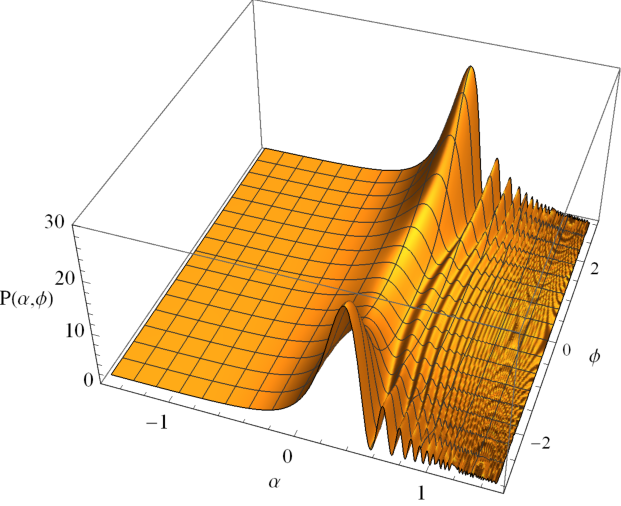}
\caption{\label{fig:model}The probability $P(\alpha,\phi) = |\psi[\alpha,\phi]|^{2}$ in the slow-roll limit ($V_{0} = 0.075$, $m = 0.5$, and $\Delta = 3$).}
\end{center}
\end{figure}

\begin{figure}
\begin{center}
\includegraphics[scale=1]{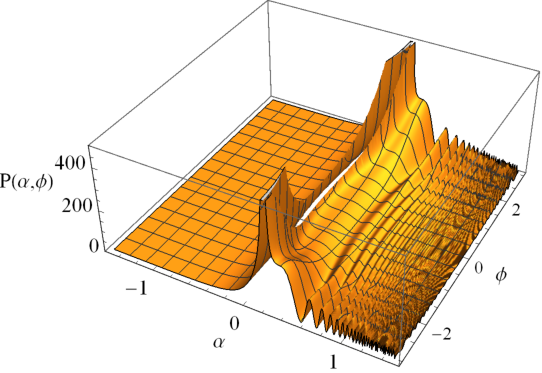}
\caption{\label{fig:model_anal}The analytic form of the probability $P(\alpha,\phi) = |\psi[\alpha,\phi]|^{2}$ in the slow-roll limit ($V_{0} = 0.075$, $m = 0.5$, and $\Delta = 3$).}
\end{center}
\end{figure}

A numerical demonstration is shown in Fig.~\ref{fig:model} where we chose a parameter set $V_{0} = 0.075$, $m = 0.5$, and $\Delta = 3$ which is within the slow-roll limit. 
The boundary condition Eq.~(\ref{eq:nb}) was set at $\alpha_{0} = -3$ which satisfies $a \ll \mathcal{H}^{-1}$.
The typical behavior of the numerical solution along the $\alpha$ axis is that, as analytically expected, there is the first dominant peak near $\alpha \simeq 0.4$ which corresponds to the horizon for the given set of parameters.
For $\alpha >  \ln (\mathcal{H}^{-1})$ corresponding to super-horizon scales, there appears small oscillations as we expect from the analytic function Eq.~(\ref{eq:nb2}).
Comparing to those analytic approximations of Eqs.~(\ref{eq:nb}) and (\ref{eq:nb2}) for sub- and super-horizon scales, respectively (Fig.~\ref{fig:model_anal}), we see that our numerical result consistently describes the nature of analytic solutions of the wave function except the divergent artifact of analytic approximations at the transition region $a \sim \mathcal{H}^{-1}$.


\subsection{Probabilities}

\begin{figure}
\begin{center}
\includegraphics[scale=0.8]{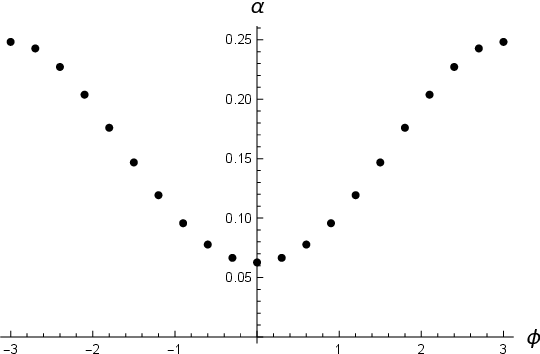}
\includegraphics[scale=0.8]{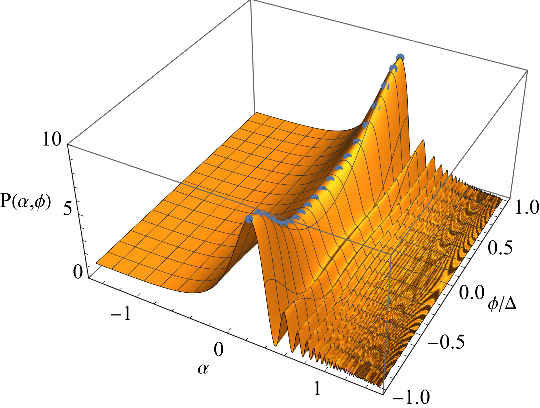}
\caption{\label{fig:alpha1}$\alpha$ (left) and $P(\alpha,\phi)$ (right) along the first peak ($V_{0} = 0.1$, $m = 0.5$, and $\Delta = 3$).}
\end{center}
\end{figure}

A proper interpretation of our numerical solution requires more details of the most dominant peak, e.g., the first steepest-descent.
Especially, we are interested in the probability. Fig.~\ref{fig:alpha1} shows the detailed analysis along the first steepest-descent for $V_{0} = 0.1$, $m = 1$, and $\Delta = 3$ case. 
In the left panel of the figure, $\alpha$ was depicted as a function of $\phi$ along the first dominant peak. 
As a check, in the right panel, one can see that the same points are lying on the first dominant peak.

It is worthwhile to compare with the analytic expectation $P = A \exp \left( 3/ 8V \right)$ with $A$ is a constant.
In order to check this, we define the measured quantity
\begin{eqnarray}
\eta \equiv \frac{3}{8 V (\log P - \log A)}.
\end{eqnarray}
If the analytic formula is a good approximation, $\eta$ from the numerical solution would be close to one, otherwise deviates.
The ambiguity of the overall normalization, i.e., $A$-dependence can be removed by setting $\eta_{\rm num} = 1$ at a certain $\phi$, e.g., the local minimum.
Then the deviation at the local maximum of the potential will be a good measure for comparison to the analytic solution.
\begin{figure}
\begin{center}
\includegraphics[scale=1]{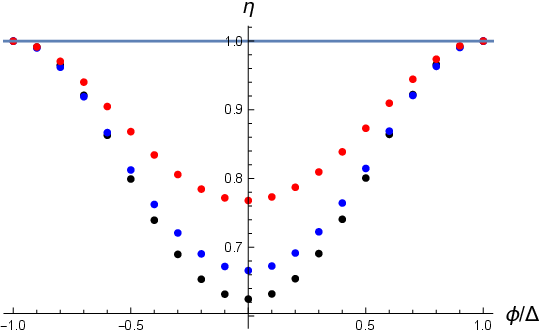}
\caption{\label{fig:probs}$\eta$ by varying $V_{0} = 0.1$ (black), $0.2$ (blue), and $0.5$ (red) with $m = 1$ and $\Delta = 3$, where the blue line is $\eta = 1$.}
\end{center}
\end{figure}

Followings are physically essential points of the probabilities.
\begin{itemize}
\item[-- 1.] \textit{Analytic limit}: Interestingly, as $V_{0}$ increases (equivalently, as the potential becomes flat more and more, satisfying the slow-roll condition), the measure $\eta$ approaches to one (Fig.~\ref{fig:probs}). 
This shows that the analytic approximation is very good for ultra-slow-roll cases.
Therefore, we confirm that this numerical approach is consistent with the analytic expectations.

\item[-- 2.] \textit{Beyond Hawking-Moss instantons}: On the other hand, as the slow-roll parameter increases, the bias from the analytic expectation becomes clearer. It is interesting to observe that, as the potential becomes steeper and steeper, the hierarchy between the local minimum and the local maximum becomes smaller than the naive expectation
\begin{eqnarray}
\log \frac{P_{\mathrm{max}}}{P_{\mathrm{min}}} \simeq \frac{3}{8} \left( \frac{1}{V_{\mathrm{max}}} - \frac{1}{V_{\mathrm{min}}} \right),
\end{eqnarray}
which is known as the Hawking-Moss instantons \cite{Hawking:1981fz}, where $P_{\mathrm{max}}$ and $P_{\mathrm{min}}$ denote the probability at the local maximum and the local minimum, respectively. This shows that the Hawking-Moss instanton is just an approximate description and the real tunneling (thermal excitation) process depends not only on the initial and final conditions but also on the entire field space (see also \cite{Weinberg:2006pc}).
\end{itemize}

About the latter point, i.e., a bias of $\eta$ from exact $1$, one may understand by assuming that $A$ is $\phi$-dependent. In this case,
\begin{eqnarray}
A(\phi) e^{-S_{\mathrm{E}}(\phi)} = e^{-S_{\mathrm{E}}(\phi) + \log A(\phi)} = e^{-S'(\phi)},
\end{eqnarray}
and hence, the Euclidean action $S_{\mathrm{E}}(\phi)$ can be modified due to the $\phi$-dependence of $A$. This might be a partial explanation for results beyond the Hawking-Moss instanton approximation.

\section{\label{sec:con}Conclusion}

In this paper, we investigated the Hartle-Hawking wave function by numerically solving the Wheeler-DeWitt equation. Historically, our paper is the \textit{first trial} to see the total wave function without using the WKB approximation. In the slow-roll limit, the result has turned out to be consistent with the analytic approximations. There exists the first steepest-descent, and the probability distribution is consistent with the instanton computations, and, as the potential becomes flatter and flatter, falling into an ultra-slow-roll regime, the difference between analytic expectations and numerical results becomes smaller and smaller. However, it is fair to say that there is little difference from the analytic results. This is an interesting cross-check of the instanton approximation. However, this approach is worthwhile compared to the previous work, because we did not rely on approximation, e.g., the ultra slow-roll approximation with the WKB approximation \cite{Vilenkin:1987kf,Hartle:1983ai} nor the instanton approximation \cite{Hartle:2008ng}; rather, we solved the Wheeler-DeWitt equation directly using numerical methods.

For a clear definite interpretation over all regions of $(\alpha, \phi)$ space, clarifications of some concepts are necessary but it is beyond the scope of this paper. One of the most important topics is the \textit{decoherence}. In the classical domain, if a classical universe is created, there must be a connection with the \textit{decoherence}. In other words, the decoherence condition must be clarified for the parameters in the classical regime.

It is worthwhile to compare with the classicality condition of the Euclidean path integral approach \cite{Hartle:2008ng,Yeom:2021twr}. In our computations, we could not find any evidence to exclude initial conditions that violate the classicality condition. However, as the potential shape approaches the non-slow-roll regime, a new phenomenon may appear in the large $\alpha$ limit. We leave this topic for a future investigation.

As the shape of the potential changes, numerical results deviate from the analytic expectations.
This is not surprising since analytic results as approximations are expected to have a validity limit, and instanton approximations are still relevant but their applicability is limited. 
In this sense, it will be very worthwhile to compute the probabilities and compare them with Euclidean computations not only with the local minimum but also with (steep) local maximum \cite{Hwang:2011mp}.

Finally, it will also be worthwhile to solve the Wheeler-DeWitt equation for non-periodic models instead of the periodic potential we considered here.
The periodic potential can be motivated from axions \cite{Hwang:2012bd}, and applications for realistic inflationary cosmology will be an interesting future research topic \cite{Hwang:2013nja}. 
Especially, applications for multi-field inflation models \cite{Hwang:2014vba} might be an interesting future research direction. 
We will touch on these topics in future works.

\newpage

\section*{Acknowledgments}

DY is supported by the National Research Foundation of Korea (NRF) grant funded by the Korean government: 2021R1C1C1008622, 2021R1A4A5031460. SK is supported by the Korea Initiative for fostering University of Research and Innovation Program of the National Research Foundation of Korea (2020M3H1A1077095). WP is supported by the Research Base Construction Fund Support Program funded by Jeonbuk National University in 2022, the Basic Science Research Program through the National Research Foundation of Korea funded by the Ministry of Education Grant No. 2017R1D1A1B06035959, and the National Research Foundation of Korea (NRF) Grant No. 2022R1A4A503036211.

\end{document}